\newif\ifjournal\journalfalse

\documentclass[aip,pop,reprint,amsmath,amssymb,groupedaddress]{revtex4-1}

\usepackage{graphicx}
\usepackage{bm}
\renewcommand{\vec}[1]{\boldsymbol{#1}}

\newcommand{\grad}{\nabla}
\renewcommand{\div}{\nabla \cdot}
\newcommand{\rot}{\nabla \times}

\def\grl{{Geophys. Res. Lett.} }
\def\jgr{{J. Geophys. Res.} }
\def\ang{{Ann. Geophys.} }
\def\prl{{Phys. Rev. Lett.} }
\def\pop{{Phys. Plasmas} }

\begin{document}

\title{The inner structure of collisionless magnetic reconnection:\\
The electron-frame dissipation measure and Hall fields}

\author{Seiji Zenitani}
\altaffiliation{Present address: National Astronomical Observatory of Japan, 2-21-1 Osawa, Mitaka, Tokyo 181-8588, Japan. Electric mail: seiji.zenitani@nao.ac.jp.}
\author{Michael Hesse}
\author{Alex Klimas}
\author{Carrie Black}
\author{Masha Kuznetsova}
\affiliation{NASA Goddard Space Flight Center, Greenbelt, Maryland 20771, USA}

\ifjournal
\else
\date{submitted 31 August 2011; accepted 21 October 2011; published 14 December 2011}
\fi

\begin{abstract}
It was recently proposed that
the electron-frame dissipation measure,
the energy transfer from the electromagnetic field to plasmas in the electron's rest frame,
identifies the dissipation region of collisionless magnetic reconnection
[Zenitani {\it et al.} Phys. Rev. Lett. {\bf 106}, 195003 (2011)].
The measure is further applied to the electron-scale structures of antiparallel reconnection,
by using two-dimensional particle-in-cell simulations. 
The size of the central dissipation region is controlled by the electron-ion mass ratio,
suggesting that electron physics is essential.
A narrow electron jet extends along the outflow direction
until it reaches an electron shock.
The jet region appears to be anti-dissipative. 
At the shock, electron heating is relevant to a magnetic cavity signature.
The results are summarized to a unified picture of
the single dissipation region in a Hall magnetic geometry.
\end{abstract}


\maketitle

\section{Introduction}

Collisionless magnetic reconnection drives explosive events
in space plasma environments.
On the large scale,
reconnection is an MHD-scale process but is facilitated
by a compact ``diffusion region'' surrounding the reconnection point,
where kinetic physics plays a role.
In the Hall reconnection model,\citep{birn01,shay01,drake07}
it is thought that the electron diffusion region (EDR),
where the electrons decouple from field lines,
is embedded in the ion's diffusion region \citep{drake07}. 
The EDR is related to a localized dissipation region
that allows fast reconnection, and
it also adjusts the overall system evolution to the outer ion physics.\citep{birn01,shay01}
Owing to its importance in reconnection,
the EDR is one of the most important targets of
NASA's upcoming Magnetospheric MultiScale (MMS) mission
({\tt http://mms.space.swri.edu/}),
which will be the first to probe the electron-scale physics.

The Hall reconnection model has been recently challenged
by kinetic particle-in-cell (PIC) simulations
that sufficiently resolve electron-scale structures\cite{dau06,keizo06,kari07,shay07}.
It has been argued that
the EDR is no longer localized but
it unexpectedly stretches in the outflow directions\cite{dau06,keizo06}. 
Furthermore, it appears to embody a two-scale structure of
inner and outer regions \cite{kari07,shay07}.
Minor differences aside,
the inner region is a compact region
containing the reconnection site.
It features
a strong out-of-plane electron current
and
a dissipative electric field
$\vec{E}' = \vec{E} + \vec{v}_e \times \vec{B} \ne 0$
to transport the magnetic fields. 
The outer region extends in the outflow direction,
accompanied by a fast electron jet,
often denoted as the ``super-Alfv\'{e}nic'' jet.
Since the jet outruns the moving magnetic field\citep{prit01a},
the out-of-plane component of $\vec{E}'$ has
the opposite polarity from that in the inner region. 
Satellite observations corroborate these
structures in near-Earth reconnection sites\citep{phan07,nagai11}.

The role of the EDR is a subject of recent debate. 
In principle, many works have agreed
that the reconnection rate is controlled by the inner region or
the inner region adjusts its size to the global reconnection rate
\citep{dau06,shay07,drake08,klimas08,wan08d,klimas10}.
Its long-term, time-dependent behavior remains unclear. 
\citet{dau06} argued that
the (inner) EDR stretches in the outflow direction.
They further proposed an idea that
the secondary island formation maintains fast reconnection
by cutting and shortening the elongated EDR. 
In contrast, \citet{shay07} demonstrated
that the inner region remains at the finite length
and
that the reconnection is fast, quasi-steady, and laminar. 
Regarding the outer region, 
it does not appear to constrain the reconnection rate
\citep{shay07,drake08,klimas08,wan08d},
because reconnection remains fast
while the electron jet extends a large distance.
\citet{hesse08} found that
the electron flow essentially consists of
the ${\mathbf E}\times{\mathbf B}$ convection and
the diamagnetic current for the field reversal. 
Thus, even though it is frequently mentioned by
the outer EDR \citep{kari07,shay07},
it is not clear whether the outer region is dissipative.
Furthermore, it is not well understood
how the electron jet is terminated. 

Recently, \citet{zeni11c} proposed a new measure
to identify the dissipation region of reconnection. 
Considering the energy dissipation in the rest frame of electron's bulk motion,
they introduced the {\it electron-frame dissipation measure},
\begin{eqnarray}
\label{eq:EDR}
D_e =
\gamma_e \big[
\vec{j} \cdot (\vec{E}+\vec{v}_e\times\vec{B}) - \rho_c ( \vec{v}_e \cdot \vec{E} )
\big],
\end{eqnarray}
where
$\gamma_e=[1-(v_e/c)^2]^{-1/2}$ is the Lorentz factor
for the electron velocity
and $\rho_c$ is the charge density.
This measure is a Lorentz-invariant, and
it is related to the nonideal energy conversion,
which is essential for reconnection problems.
Their PIC simulations demonstrated that
$D_e$ identifies a single dissipation region
surrounding the reconnection site.
This is substantially different from the previous two-scale picture.

The purpose of this paper is to
organize our knowledge on
the dissipation region
in collisionless magnetic reconnection
in line of recent theoretical progress \citep{hesse08,zeni11c}.
Using high-resolution two-dimensional PIC simulations,
we perform a detailed investigation of internal structures
in basic antiparallel configurations.
In particular, the previous two-scale EDR are reexamined
by using the electron-frame dissipation measure $D_e$.
The structures are better understood in a Hall magnetic geometry.
We further explore the termination region of the electron jet.

This paper is organized as follows.
We briefly review the dissipation measure $D_e$ in Sec. \ref{sec:theory}.
We describe our numerical setup in Sec. \ref{sec:setup}.
We present the simulation results in Sec. \ref{sec:results}.
We carefully investigate fine structures in the following subsections.
Section \ref{sec:discussion} contains discussion and summary.

\section{The electron-frame dissipation measure}
\label{sec:theory}

For better understanding, we review
a basic concept in Ref.~\onlinecite{zeni11c}
within the nonrelativistic physics.
The essence is
to evaluate the Ohmic dissipation
in the rest frame of electron's bulk motion.
Let the prime sign ($'$) denote quantities in the electron frame. 
A simple algebra yields
\begin{eqnarray}
\label{eq:j'}
\vec{j}'
&=&
e n_i \vec{v}'_i
= e (n_i \vec{v}_i - n_e \vec{v}_e) - e (n_i-n_e) \vec{v}_e
\nonumber \\
&=&
\vec{j} - \rho_c \vec{v}_e
.
\end{eqnarray}
{
The last term comes from the convection current ($\rho_c \vec{v}_e$),
arising from the motion of the charged frame.}
The electric field in the electron frame
is given by the nonideal electric field,
\begin{eqnarray}
\label{eq:E'}
\vec{E}' = \vec{E} + \vec{v}_e \times \vec{B}.
\end{eqnarray}
The energy dissipation in the electron frame yields
\begin{eqnarray}
\label{eq:EDR1}
D_e
&=&
\vec{j}'\cdot\vec{E}' =
\vec{j} \cdot (\vec{E} + \vec{v}_e\times\vec{B})
- \rho_c \vec{v}_e \cdot \vec{E}
.
\end{eqnarray}
This is an invariant scalar
with respect to the Galilean transformation
and
is equivalent to Eq. \ref{eq:EDR}
in the limit of $\gamma_e \rightarrow 1$.
We employ this nonrelativistic formula throughout this paper.
We often discuss the composition of $D_e$
in the following form,
\begin{eqnarray}
\label{eq:EDR2}
D_e
&=&
j_x E'_x + j_y E'_y + j_z E'_z
- \rho_c \vec{v}_e \cdot \vec{E}
.
\end{eqnarray}
The last term deals with the effective energy conversion by the convection current.
Hereafter we refer to it by the ``charge term.''

In a neutral plasma, one can obtain
$D_e=\vec{j}\cdot\vec{E}'$ by dropping the charge term.
This reduced form is equivalent to
the Joule dissipation in the plasma rest frame in \citet{birn05b,birn09},
who discussed the energy transfer in MHD and kinetic reconnection systems.

\section{Numerical method}
\label{sec:setup}

We use a partially-implicit PIC code \citep{hesse99,hesse08,zeni11c}.
The length, time, and velocity are normalized by
the ion inertial length $d_i=c/\omega_{pi}$,
the ion cyclotron time $\Omega_{ci}^{-1}=m_i/(eB_0)$, and
the ion Alfv\'{e}n speed $c_{Ai}=B_0/(\mu_0 m_i n_0)^{1/2}$, respectively.
Here, $\omega_{pi}=( e^2n_0/\epsilon_0 m_i)^{1/2}$ is the ion plasma frequency,
$n_0$ is the reference density, and
$B_0$ is the reference magnetic field.
We employ a Harris-like configuration,
$\vec{B}(z)=B_0 \tanh(z/L) \vec{\hat{x}}$ and
$n(z) = n_{0} [0.2 + \cosh^{-2}(z/L)]$,
where $L=0.5 d_i$ is the half thickness of the current sheet.
The electron-ion temperature ratio is $T_e/T_i=0.2$.
Periodic ($x$) and reflecting wall ($z$) boundaries are used.

In the main run (run 1A),
the mass ratio is $m_i/m_e=100$.
The ratio of the electron plasma frequency to the electron cyclotron frequency
is $\omega_{pe}/\Omega_{ce}=4$.
In this case,
the speed of light is $(c/c_{Ai})=(m_i/m_e)^{1/2}(\omega_{pe}/\Omega_{ce})=40$.
The domain of $[0,76.8]\times[-19.2, 19.2]$ is resolved by
$2400 \times 1600$ cells, and
$2.2{\times}10^9$ particles are used.
We carry out several other runs,
changing the two key parameters:
the mass ratio $m_i/m_e$ and the $\omega_{pe}/\Omega_{ce}$ parameter.
All parameters are presented in Table.~\ref{table}.
In all cases, reconnection is triggered by a small flux perturbation.
$\delta A_y = - 2L B_1 \exp[-(x^2+z^2)/(2L)^2]$,
where $B_1=0.1 B_0$ is the typical amplitude of the perturbed fields.
The initial current is set up accordingly.

\begin{table}[htdp]
\begin{center}
\caption{Simulation parameters}
\begin{tabular}{cccccc}
\hline
\hline
~Run~ & $m_i/m_e$ & $\omega_{pe}/\Omega_{ce}$ & Domain size & ~~~Grid cells~~~ & ~Particles~ \\
\hline
1A & 100 & 4 & $76.8 \times 38.4$ & $2400 \times 1600$ & $2.2 \times 10^9$ \\
1B & 100 & 2 & $76.8 \times 38.4$ & $2400 \times 1600$ & $2.2 \times 10^9$ \\
2A &  25 & 4 & $102.4 \times 51.2$ & $1600 \times 1600$ & $1.4 \times 10^9$ \\
2B &  25 & 2 & $102.4 \times 51.2$ & $1600 \times 1600$ & $2.6 \times 10^9$ \\
\hline
\hline
\end{tabular}
\label{table}
\end{center}
\end{table}

\section{Results}
\label{sec:results}

\subsection{Overview}

We identify the dominant reconnection site by finding
the minimum in $x$ of the magnetic flux
$\Phi(x,t)=\frac{1}{2} \int |B_x| dz$.
Then we discuss the system evolution
by using the normalized flux transfer rate or
the reconnection rate $\mathcal{R}$,
\begin{eqnarray}
\label{eq:recrate}
\mathcal{R} = -\frac{1}{c_{A,in} B_{in}}
\frac{\rm d}{\rm dt}
\Phi_{min},
\end{eqnarray}
where the subscript $in$ denotes quantities
at $3 d_i$ upstream of the dominant reconnection site. 

\begin{figure}[thp]
\begin{center}
\ifjournal
\includegraphics[width=\columnwidth]{f1.eps}
\else
\includegraphics[width=\columnwidth]{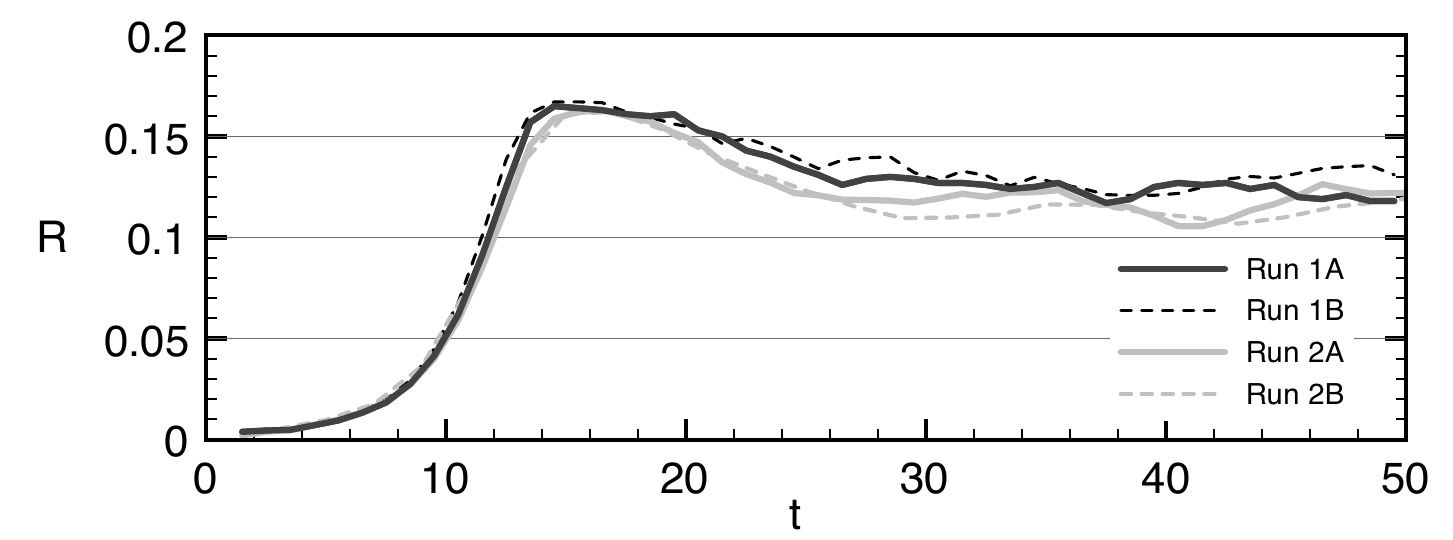}
\fi
\caption{
\label{fig:recrate}
The normalized reconnection rates $\mathcal{R}$ (Eq. \ref{eq:recrate})
as a function of a time ($\Omega_{ci}t$).
}
\end{center}
\end{figure}

Figure \ref{fig:recrate} shows the time evolution of $\mathcal{R}$.
The rate quickly increases, modestly overshoots at $t{\sim}15$,
and then approaches a quasisteady value of $0.12$--$0.13$.
Such quasisteady evolution corroborates previous investigations.\citep{shay07} 
The rates are insensitive to the mass ratio in this normalization.\citep{hesse99}
The parameter $\omega_{pe}/\Omega_{ce}$ does not
make a significant difference also.
We confirmed that
the reconnection rates are independent of domain sizes
in the timescale of our interest,
by carrying out supplemental small runs.
Meanwhile, as will be discussed later,
the periodic domain effects start to modulate
the outflow structures after $t{\sim}40$ in runs 1A and 1B.

\begin{figure}[thp]
\begin{center}
\ifjournal
\includegraphics[width=\columnwidth]{f2_new.eps}
\else
\includegraphics[width=\columnwidth]{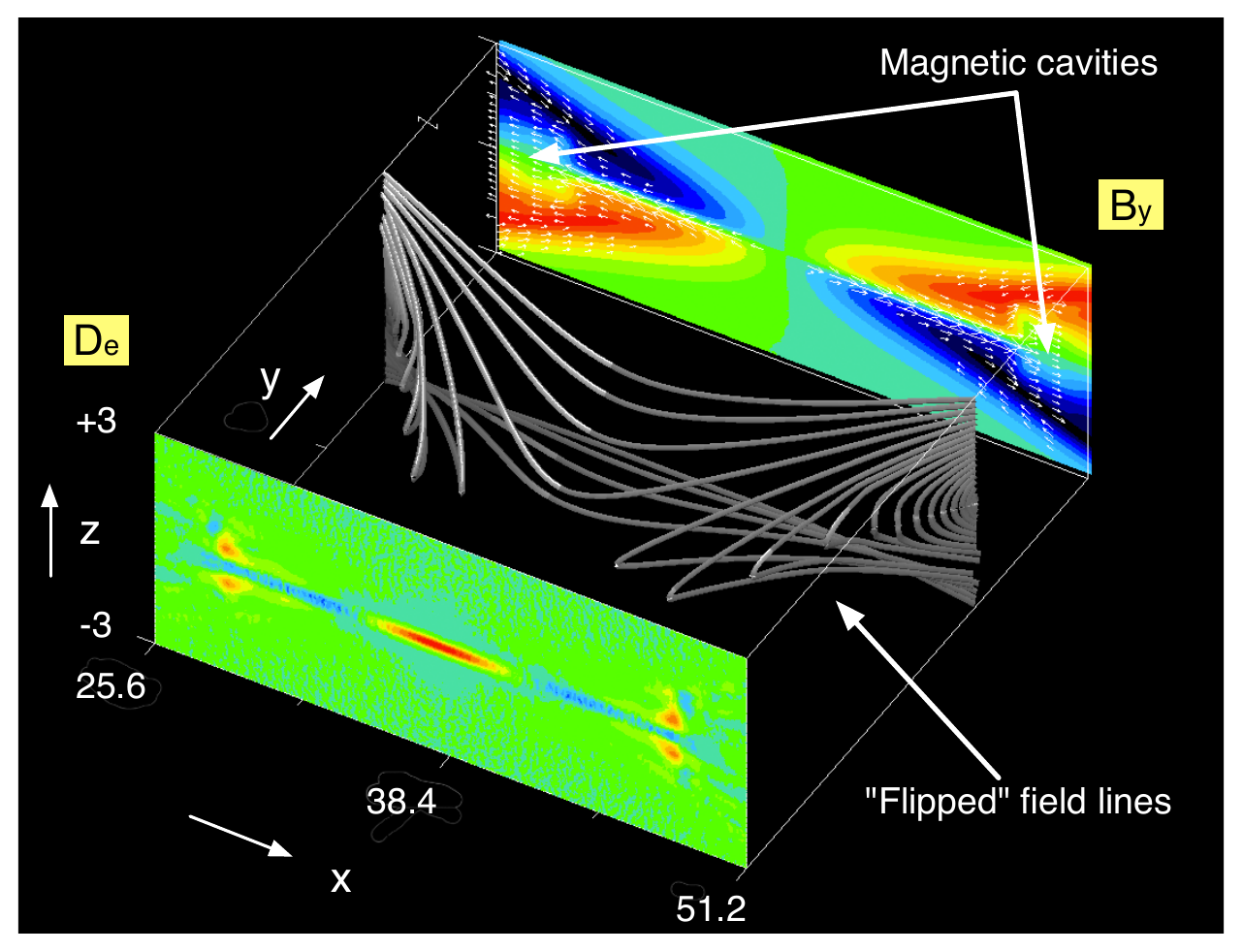}
\fi
\caption{(Color online)
\label{fig:3D}
Magnetic field line structure in run 1A, averaged over $t=35$-$36$.
Rear panel: the out-of-plane magnetic field $B_y$.
Front panel: the electron-frame dissipation measure $D_e$
}
\end{center}
\end{figure}

Figure \ref{fig:3D} presents a snapshot of run 1A. 
Physical quantities are averaged over $t=35$--$36$.
At this stage, the reconnection is steadily going on
as can be seen in Figure \ref{fig:recrate}, and
the characteristic structures are well developed. 
This is the stage of our primary interest. 
The gray lines show magnetic field lines in 3D. 
Magnetic field lines are flipped,
because they tend to follow the electron's flow in the $-y$ direction.
This is a well-known signature of Hall reconnection.
For example, \citet{huba02} showed
a similar field-line structure by using Hall MHD simulations.
For example, \citet{yamada06} found a similar structure in an experiment device.
Our fully-kinetic simulation is consistent with these results. 
The rear panel in Figure \ref{fig:3D} shows the out-of-plane magnetic field $B_y$.
The color indicates the polarity: $B_y>0$ in red and $B_y<0$ in blue.
The projection of flipped field lines leads to
a characteristic ``quadrupole'' pattern\citep{sonnerup79,terasawa83} in $B_y$. 
In addition, as indicated in Figure \ref{fig:3D},
there are two ``magnetic cavities'' in the outflow region,
where $B_y$ becomes weaker.
This signature is often found in PIC simulations of Hall reconnection,
but this has never been discussed before. 
The front panel shows the electron-frame dissipation measure $D_e$ (Eq.~\ref{eq:EDR1}).
A central red region of $D_e>0$ is
the dissipation region\citep{zeni11c}.
We will investigate this region in detail later in this paper.

\begin{figure*}[thp]
\begin{center}
\ifjournal
\includegraphics[width=\textwidth]{f3.eps}
\else
\includegraphics[width=\textwidth]{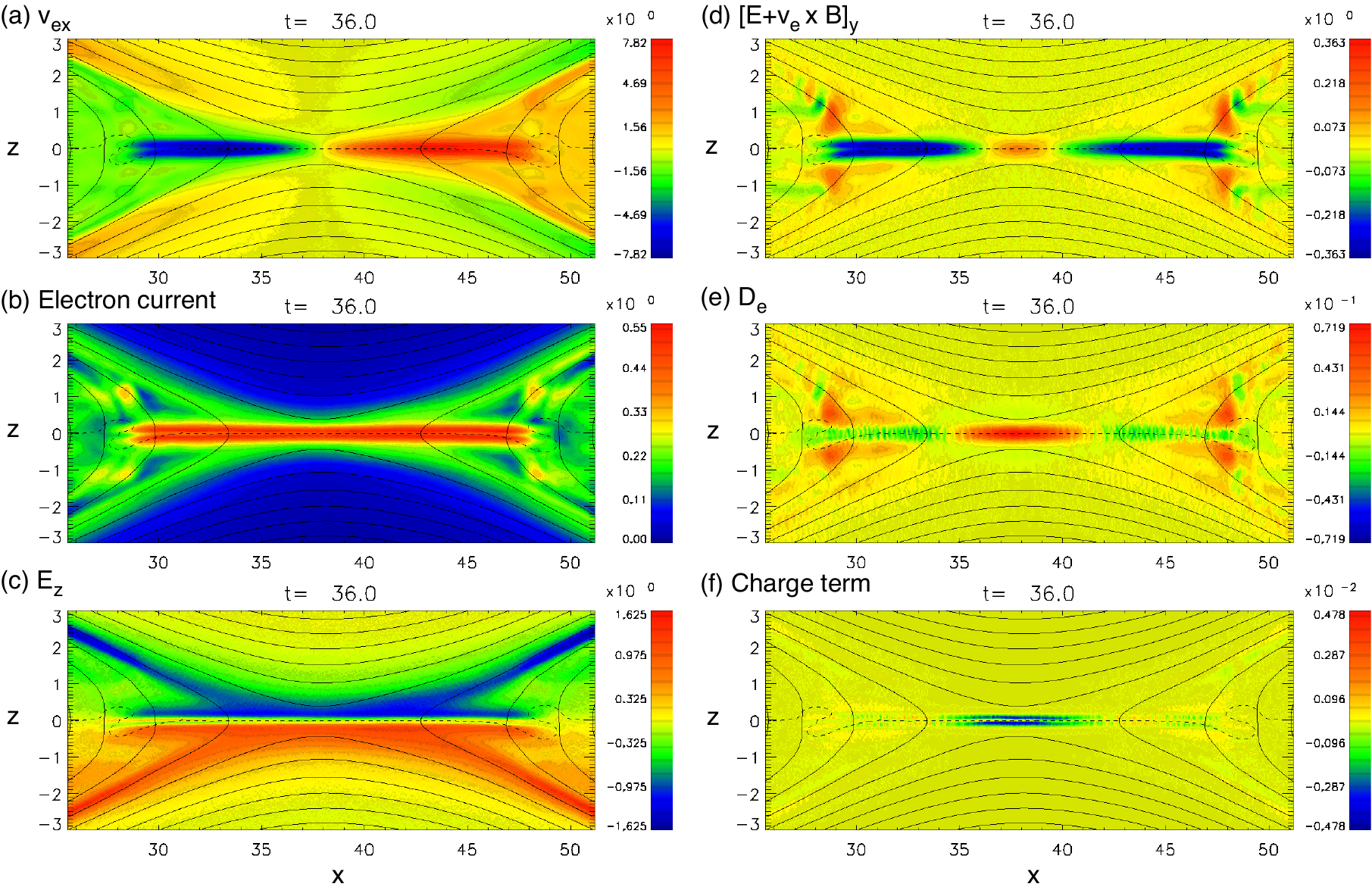}
\fi
\caption{(Color online)
\label{fig:35}
Averaged profile of run 1A over $t=35$--$36$.
(a) The electron outflow velocity $v_{ex}$,
(b) the electron current density $|j_e|$ in units of $J_0$,
(c) the electric field $E_z$ in units of $c_{Ai}B_0$,
(d) the out-of-plane component of the electron nonideal condition
$(\vec{E}+\vec{v}_{e}\times\vec{B})_y$ in units of $c_{Ai}B_0$,
(e) the electron-frame dissipation measure $D_e$ in units of $c_{Ai}B_0J_0$,
and
(f) the charge term ($-\rho_c \vec{v}_e\cdot\vec{E}$) in $D_e$.
The contour lines are in-plane magnetic field lines.
The dash line indicates the field reversal, $B_x=0$.
}
\end{center}
\end{figure*}

Figure \ref{fig:35} show various averaged quantities over $t=35$-$36$ in run 1A. 
Figure \ref{fig:35}(a) shows the electron outflow velocity $v_{ex}$.
One can see bi-directional jets from the reconnection site.
They travel much faster than 
the upstream Alfv\'{e}n speed $(0.2)^{-1/2} c_{Ai} \sim 2.24$,
which approximates the outflow speed in an MHD scale.
The electron jets are remarkably narrow.
Their widths are on an order of the local electron inertial length.
Outside the separatrices,
there are weak reverse flows toward the reconnection site. 
In Figure \ref{fig:35}(b) we show the electron current density $|j_e|$.
It is distinctly strong in a narrow region near the neutral plane ($z=0$).
We note that it is remarkably flat in the $x$-direction. 
Thus, $|j_e|$ appears to be a good marker of the interesting region.
Hereafter we call the region an ``electron current layer.''

Figure \ref{fig:35}(c) shows the vertical electric field $E_z$.
Near the reconnection site,
one can see a bipolar signature across the electron current layer:
$E_z$ is negative for $z>0$ and positive for $z<0$.
This is called the Hall or polarization electric field
\citep{shay98b,arz01,prit01b,hoshino01,keizo06}.
This is supported by charge separation between
a broader distribution of meandering ions and
a narrow distribution of meandering electrons. 
The bipolar peaks $|E_{z}|$ are typically $8$ times stronger than
the reconnection electric field $E_y$.
In the inflow region, outside the electron current layer,
magnetized electrons tend to travel in the -$y$-direction\citep{ishizawa04}. 
In the outflow region, the $E_z$ region consists of
two parts:
the large-scale X-shaped structure along the separatrices and
the top/bottom boundaries of the electron current layer\citep{hoshino01}.
We think the latter is related to the electron meandering motion,
similarly as the central region. 
In addition, as mentioned by a previous work\citep{chen11},
one can recognize an inverted signature of $E_z$
in a very vicinity of the neutral plane ($|z|<0.1$)
inside the electron current layer. 
For example, $E_z$ is positive for $z>0$ and negative for $z<0$
around $x \sim 29$ and $x\sim 47$.
These ``inversion electric field''\citep{chen11} are supported by
charge separation inside the electron meandering orbit.

Shown in Figure \ref{fig:35}(d) is the out-of-plane component of
the electron Ohm's law, $E'_y=[\vec{E}+\vec{v}_e\times\vec{B}]_y$,
in unit of $c_{Ai}B_0$.
One can recognize two distinct regions along the electron jets.
One is the compact region surrounding the reconnection site,
where $E'_y$ is positive.
The other regions extend in the outflow direction and $E'_y$ is negative,
because the fast electron jets outrun the field convection.
Those two are often referred as
the inner and outer EDRs in recent literature \citep{kari07,shay07}. 
Note that $E'_y \ne 0$ just tells us that
the ideal assumption breaks down.
It does not always indicate magnetic dissipation.

Figure \ref{fig:35}(e) presents
the electron-frame dissipation measure $D_e$
(Eq.~\ref{eq:EDR1})
in units of $c_{Ai}B_0 J_0$,
where $J_0$ is the initial Harris current density. 
Note that the color scale is different from a reduced color scale in Fig. \ref{fig:3D}. 
One can see that $D_e$ gives a different picture from Figure \ref{fig:35}(d).
The dissipation region with $D_e>0$ is located near the reconnection point
[the red region near $35<x<41$ in Fig. \ref{fig:35}(e)].
On the other hand,
$D_e$ is weakly negative in the outflow regions,
as can be seen in light green in the Figure.
There, its amplitude $|D_e|$ is substantially smaller than near the reconnection point. 
Figure \ref{fig:35}(f) presents the last charge term in Eq. \ref{eq:EDR2}
in the same unit as Fig. \ref{fig:35}(d).
This term appears only in the close vicinity of
the above central region of $D_e>0$, and
it is an order of magnitude smaller than $D_e$.

\subsection{Dissipation region}
\label{sec:DR}

\begin{figure}[thp]
\begin{center}
\ifjournal
\includegraphics[width=\columnwidth]{f4.eps}
\else
\includegraphics[width=\columnwidth]{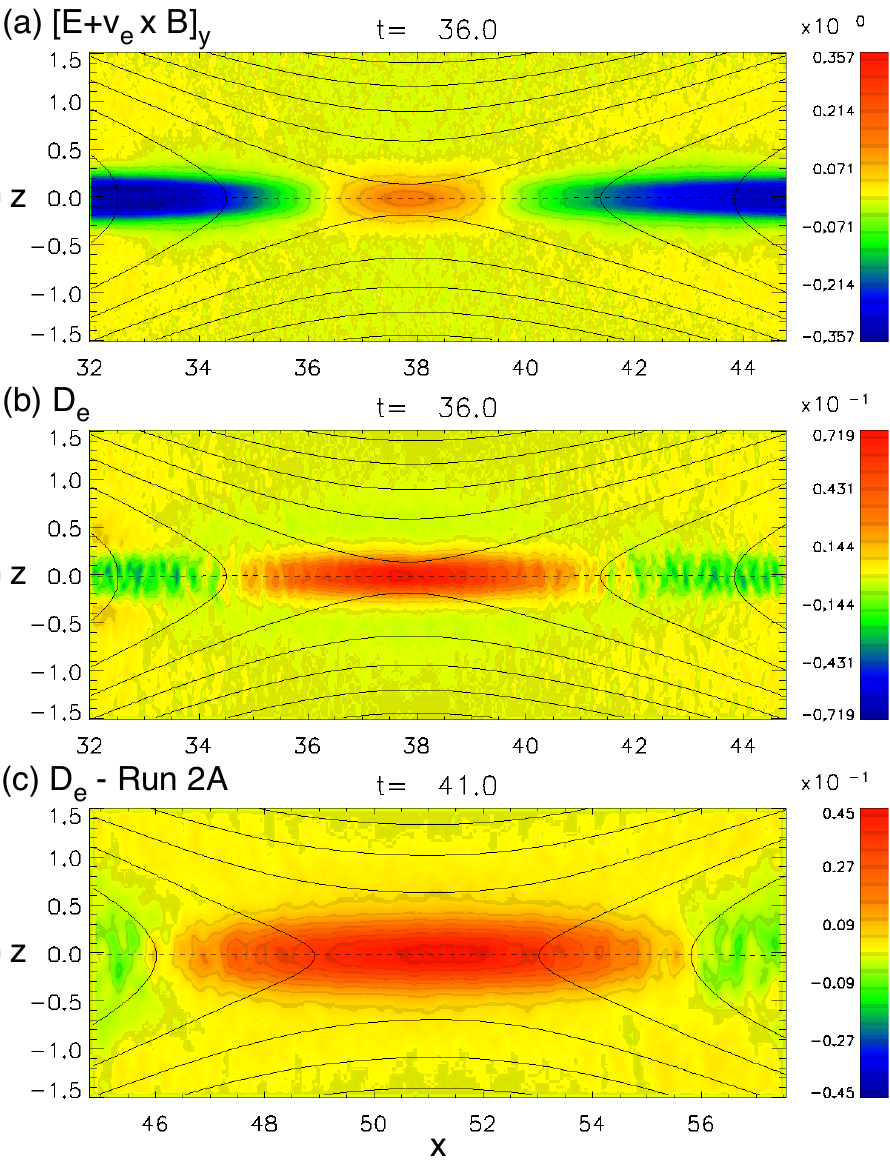}
\fi
\caption{(Color online)
\label{fig:DR}
(a) the out-of-plane component of the electron nonideal condition
$(\vec{E}+\vec{v}_{e}\times\vec{B})_y$ around the reconnection site,
(b) the electron-frame dissipation measure $D_e$, and
(c) $D_e$ in run 2A.
}
\end{center}
\end{figure}

Here, we focus on the dissipation region.
The panels in Figure \ref{fig:DR} present
the electron nonidealness $E'_y$, 
the dissipation measure $D_e$,
and
the dissipation measure $D_e$ in run 2A
at lower mass ratio of $m_i/m_e=25$. 
We find that the lengths of the electron current layer and
the dissipation region are nearly stationary in the later phase of simulations.
The panels in Figure \ref{fig:DR} are taken from the stationary phase.

Comparing Figs. \ref{fig:DR}(a) and (b),
one can see that the dissipation region is longer in the outflow direction
than the inner region defined by $E'_y$\citep{shay07,klimas08}. 
The comparison of Figs. \ref{fig:DR}(b) and (c) suggests that
the size of the dissipation region is controlled by the electron mass. 
In our ion-based units,
the electron inertial length scales with $(m_e/m_i)^{1/2}$ and
the electron bounce width scales with $(m_e/m_i)^{1/4}$
(Refs.~\onlinecite{hesse99}, \onlinecite{biskamp71}).
We find that the length and the width of the dissipation region
in run 1A is smaller than those in run 2A by a factor of $1.5$-$1.8$.
This is intermediate between $(100/25)^{1/4}$ and $(100/25)^{1/2}$.
The scaling to the mass ratio needs further investigation, but
electron physics appears to be essential for the dissipation region.

\begin{figure}[thp]
\begin{center}
\ifjournal
\includegraphics[width=\columnwidth]{f5.eps}
\else
\includegraphics[width=\columnwidth]{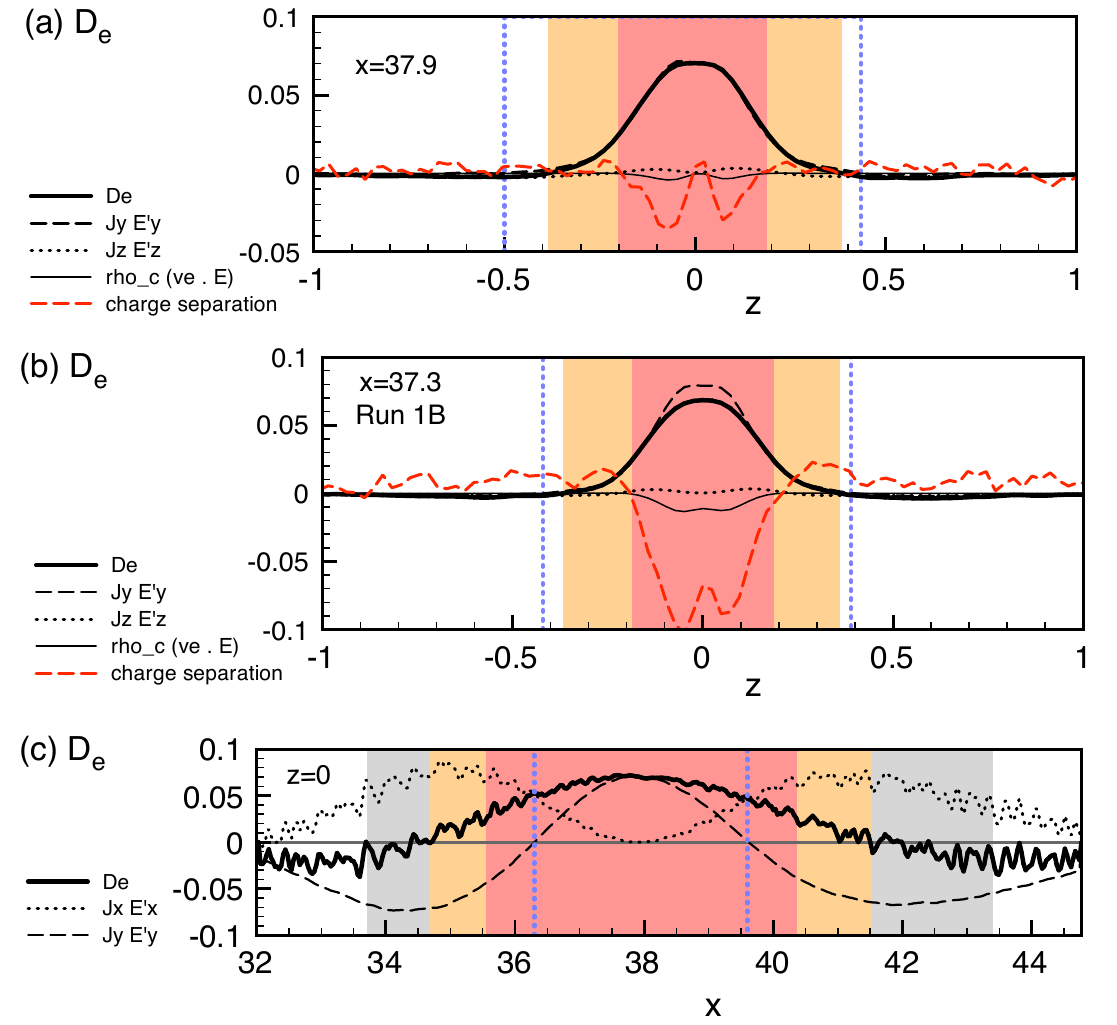}
\fi
\caption{(Color online)
\label{fig:De}
Composition of the dissipation measure $D_e$
(a) along the inflow line ($x=37.9$) in run 1A,
(b) along the inflow line ($x=37.3$) in run 1B, and
(c) along the outflow line ($z=0$) in run 1A,
}
\end{center}
\end{figure}

Next, we discuss the composition of the dissipation measure $D_e$.
Shown in Figure \ref{fig:De} are contributions from the terms in
Eq. \ref{eq:EDR2} and the charge separation
$(n_i-n_e)/(n_i+n_e)$ along the inflow and outflow lines.
The shaded regions indicate the dissipation region
by the sign ($D_e>0$; orange) and
the $e$-folding length of $D_e$ (the inner red region),
respectively.
The blue dotted lines indicate the electron ideal condition,
$E'_y=[\vec{E}+\vec{v}_e\times\vec{B}]_y=0$.
The nonideal electric field $E'_y$ is positive between
the two dotted lines.

In the inflow direction in run 1A [Fig. \ref{fig:De}(a)],
$D_e$ is well localized near the neutral plane $z=0$.
Its main contributor is the $y$-term ($j_yE'_y$).
The charge term ($-\rho_c \vec{v}_e\cdot\vec{E}$) is responsible for $-4\%$ of $D_e$.
This term has a double peak structure and comes from a double-peak
charge distribution (the red dashed line),
due to the electron bounce motion.
The charge separation is up to $-4\%$ in this case.
The $z$-term ($j_zE'_z$) has also double peaks,
because both $j_z$ and $E'_z$ have bipolar signatures. 
However, since $j_z$ is small, the $z$-term is a very minor contributor to $D_e$.
The $x$-term is negligible also.

Figure \ref{fig:De}(b) shows the composition in run 1B at lower ratio of
$\omega_{pe}/\Omega_{ce}=2$.  In this case, the charge term
is responsible for -$15{\sim}20\%$ of $D_e$, because the charge separation is significant, $\sim 10\%$.
A lower $\omega_{pe}/\Omega_{ce}$ allows more significant charge separation.
We confirm this trend at the mass ratio $m_i/m_e=25$, too. 
In the outflow direction [Figure \ref{fig:De}(c)],
one can see that the $x$-term ($j_xE'_x$) is another contributor to $D_e$.
Both $D_e$ and $j_xE'_x$ look noisy in the outflow regions,
because $E_x$ is noisy.  The gray region is defined by the spatial location of
a maximum electron outflow velocity.

Let us compare the dissipation region with the conventional inner region\citep{kari07,shay07}.
In the inflow direction [Fig. \ref{fig:De}(a)], the inner region has been defined
by the electron nonidealness $E'\ne 0$ (the blue dotted lines).
Both orange and red dissipation regions are thinner than
the conventional inner region, bounded by the blue dotted lines [Fig. \ref{fig:De}(a)].
One case see that the $e$-folding region (red) is relevant to the electron bouncing region,
featuring the charge separation.
On the other hand, there have been several definitions in the outflow direction:
the electron outflow velocity (gray region)\citep{dau06,keizo06,kari07},
the sign of the electron nonidealness (the dotted lines)\citep{shay07,klimas08,klimas10},
or the out-of-plane current profile\citep{wan08d}.
Note that Klimas et al.\citep{klimas10} employed a different formula
$-[(1/n_ee) \div \overleftrightarrow{P_e} + (m_e/e)(\vec{v}_e\cdot\grad)\vec{v}_e)]_y > 0$,
but this is almost equivalent to the popular criteria $E'_y > 0$.

\begin{figure}[thp]
\begin{center}
\ifjournal
\includegraphics[width=\columnwidth]{f6.eps}
\else
\includegraphics[width=\columnwidth]{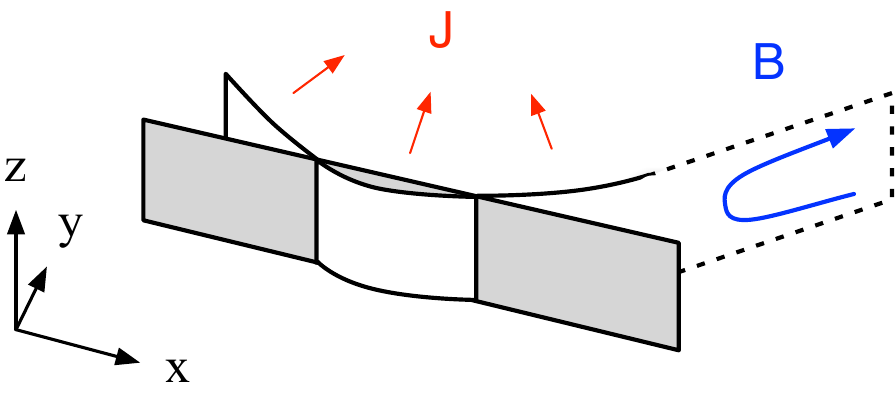}
\fi
\caption{
\label{fig:surface}
(Color online)
The Hall structure of the dissipation region.
}
\end{center}
\end{figure}

In the outflow direction,
the dissipation region is longer than
the inner region defined by $E'_y$\citep{shay07,klimas08}.
One can intuitively understand this,
considering the magnetic geometry \citep{hesse08}.
Since the field lines are flipped (Fig. \ref{fig:3D}),
we consider the curved surface
as illustrated in Figure \ref{fig:surface}.
Hereafter we call it the Hall surface.
As the field lines rotate their directions in the surface,
$\vec{j}$ ($\approx \rot \vec{B}/\mu_0$) is normal to it.
The $\vec{j}$-aligned (out-of-plane) component of $\vec{E}'$ is responsible for
the nonideal field convection in the Hall surface.
Meanwhile, at sufficiently high $\omega_{pe}/\Omega_{ce}$,
one can reduce the dissipation measure to $D_e \approx \vec{j}\cdot\vec{E}'$.
This is controlled by the $\vec{j}$-aligned component of $\vec{E}'$.
In other words, $D_e$ automatically takes care of the out-of-plane component of
$\vec{E}'$ with respect to the Hall surface.
Thus, as we move to the outflow direction, we should consider $E'_x$ as well as $E'_y$.
The positive-$D_e$ dissipation region is always longer than
the positive-$E'_y$ region\citep{shay07,klimas08},
because $D_e$ further includes contributions from $E'_x$.
According to our preliminary results (Klimas et al. in prep),
the former is commonly twice longer than the latter,
even when the reconnection region is disturbed by secondary magnetic islands.

On the other hand,
the dissipation region is shorter than
the region defined by the peak location of $v_{ex}$\citep{dau06,keizo06,kari07}.
We find that $v_{ex}$ is not very useful to define an important region
due to the following reasons.
First, it is difficult to identify the peak location,
when $v_{ex}$ gradually changes in the outflow direction. 
Second, the physical meaning of the maximum $v_{ex}$ is unclear
in the Hall geometry (Fig. \ref{fig:surface}).
The Hall surface is already rotated around there, and so,
$v_{ex}$ is highly influenced by
the out-of-plane diamagnetic flow\citep{hesse08} and
the angle of the Hall surface.

Practically, we prefer to define the dissipation region
by the $e$-folding length (red) rather than by its sign (orange). 
This is because in some cases the slope of $D_e$ is very flat around $D_e\sim 0$ in the inflow directions.
In Figures \ref{fig:De}(a) and (c), the aspect ratios of the dissipation region
are $0.39 : 4.77$ ($= 0.08: 1$) by the $e$-folding length and
$0.77 : 6.83$ ($= 0.11 : 1$) by the positive dissipation, respectively.
These values reasonably fit to the typical reconnection rate of 0.1.

\subsection{Outflow region}
\label{sec:outflow}

Next, we visit the outflow regions [$29 < x < 36$ or $41 < x < 48$; Fig. \ref{fig:35}(a)].
One can see the narrow fast electron jets\citep{kari07,shay07}.
They are flanked by slow uniform flow regions.
The region is denoted as the pedestal\citep{drake08}.
In the pedestal, the $E'_y$ component is negligible,
suggesting quasi-ideal electron flows.
There is a boundary layer with weak $E'_x$ and $E'_z$
between the jet and the pedestal.

Importantly, as shown in Figure \ref{fig:35}(e),
the dissipation measure is weakly negative $D_e < 0$ in the electron jet. 
From the viewpoint of the Hall surface (Fig. \ref{fig:surface}),
the out-of-plane component ($\sim \vec{j}$-aligned component) of $\vec{E}'$
is not exactly zero in a rotated frame \citep{hesse08}. 
The electron jet still outruns the field convection
in the outflow direction in the Hall surface.
This leads to $D_e \approx \vec{j}\cdot\vec{E}' < 0$ in the electron jet.
On the other hand,
$D_e$ is weakly positive in the pedestal and
it often has positive peaks in the boundary layers
between the jet and the pedestal.
This is visible in particular in the downstream
($29 < x < 33$ and $45 < x < 48$) and is more evident at later stages.

The fact $D_e<0$ tells us that
the electron jet region is not dissipative.
To understand this, it is useful to discuss the MHD energy balance,
as given in previous literature\citep{birn05b,birn09,zeni11c}.
We consider a mass-averaged MHD velocity in an electron-ion plasma,
\begin{eqnarray}
\label{eq:vmhd}
\vec{v}_{\rm mhd}
&=&
\frac{m_in_i\vec{v}_i+m_en_e\vec{v}_e}{m_in_i+m_en_e}.
\end{eqnarray}
The energy transfer from the fields to plasmas
in the MHD frame $D_{\rm mhd}$ is similarly given by
\begin{eqnarray}
\label{eq:dmhd0}
D_{\rm mhd}
&=& 
\vec{j} \cdot (\vec{E}+\vec{v}_{\rm mhd}\times\vec{B}) - \rho_c ( \vec{v}_{\rm mhd} \cdot \vec{E} )
.
\end{eqnarray}
A small amount of algebra leads to
\begin{eqnarray}
\label{eq:dmhd}
D_{\rm mhd}
&=& 
\frac{m_in_e+m_en_e }{m_in_i+m_en_e}~D_e 
.
\end{eqnarray}
Thus $D_e<0$ indicates $D_{\rm mhd}<0$, i.e.,
the plasma's energy is converted to the field energy even in the MHD frame. 
This contradicts with the concept of magnetic dissipation or diffusion,
which consumes the field energy in the MHD frame. 
No dissipation process takes place here.
Instead, this region is better described as ``anti-dissipative.''

\begin{figure}[thp]
\begin{center}
\ifjournal
\includegraphics[width=\columnwidth]{f7.eps}
\else
\includegraphics[width=\columnwidth]{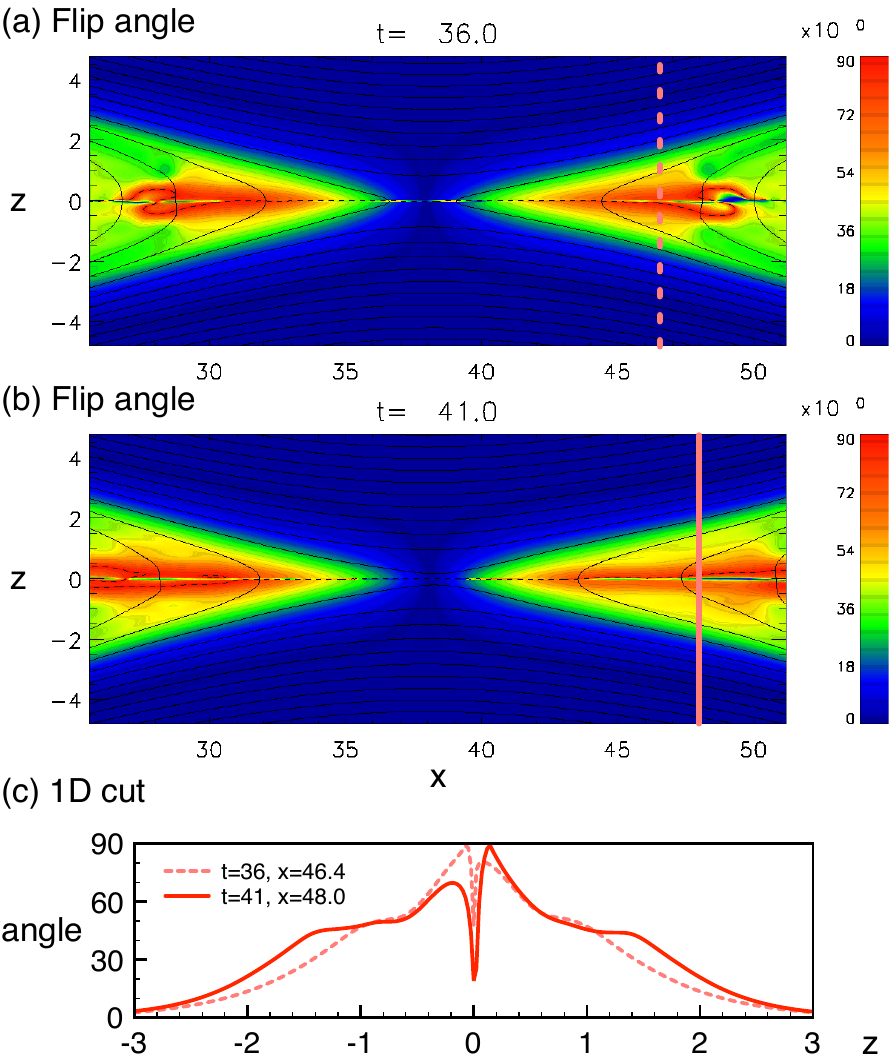}
\fi
\caption{
\label{fig:angle}
(Color online)
The flip angle [degeree] of the Hall magnetic field lines
(a) at $t=35$-$36$ and (b) at $t=40$-$41$.
(c) 1D cuts along the two lines.
}
\end{center}
\end{figure}

Figure \ref{fig:angle}(a) shows a flip angle of Hall magnetic field lines.
We calculate it by an arctangent of $B_y$ over $B_x$,
which is a good indicator outside the $z = 0$ plane.
The sign of the angle is neglected to better see the structure.
As can be seen in Figure \ref{fig:3D},
the magnetic field lines are extremely flipped.
The angle is almost $90^\circ$ in the electron jet region.
Such a strong flipping is found in recent studies
(e.g., the hodogram analysis by \citet{drake08}).
The angle changes from zero to $90^\circ$ from the upstream region to
the electron jet region.
Further observation tells us that the angle has
a two-scale structure in the same $x$-location:
a moderately flipped pedestal and
a highly flipped electron jet.
This is more evident at a later time of $t=40$-$41$ [Fig. \ref{fig:angle}(b)] and
the 1D cuts along the vertical lines [Fig. \ref{fig:angle}(c)].
In the pedestal region, the angle is $\sim 45-50^\circ$.
Note that the Hall magnetic field $|B_y|$ is usually comparable to
the antiparallel field $|B_x|$ in many simulations. 

We think the $D_e$-pattern and the field line geometry is relevant.
In the electron jet, the field lines are strongly flipped, and 
$D_e < 0$ implies that the electromagnetic fields gain energy in the electron frame.
Outside the jet, the magnetic field lines are mildly flipped and
$D_e > 0$ indicates that plasmas gain energy from the fields.
The electron-field energy transfer $\vec{j}_e \cdot \vec{E}$ shows
similar signatures as $D_e$.
We speculate that the bulk kinetic energy of the fast electron jet is
transferred from the jet to the pedestal
via the magnetic field lines that thread two regions.
This process needs to be further investigated.

\subsection{Electron shock}
\label{sec:eshock}

\begin{figure}[thp]
\begin{center}
\ifjournal
\includegraphics[width=\columnwidth]{f8.eps}
\else
\includegraphics[width=\columnwidth]{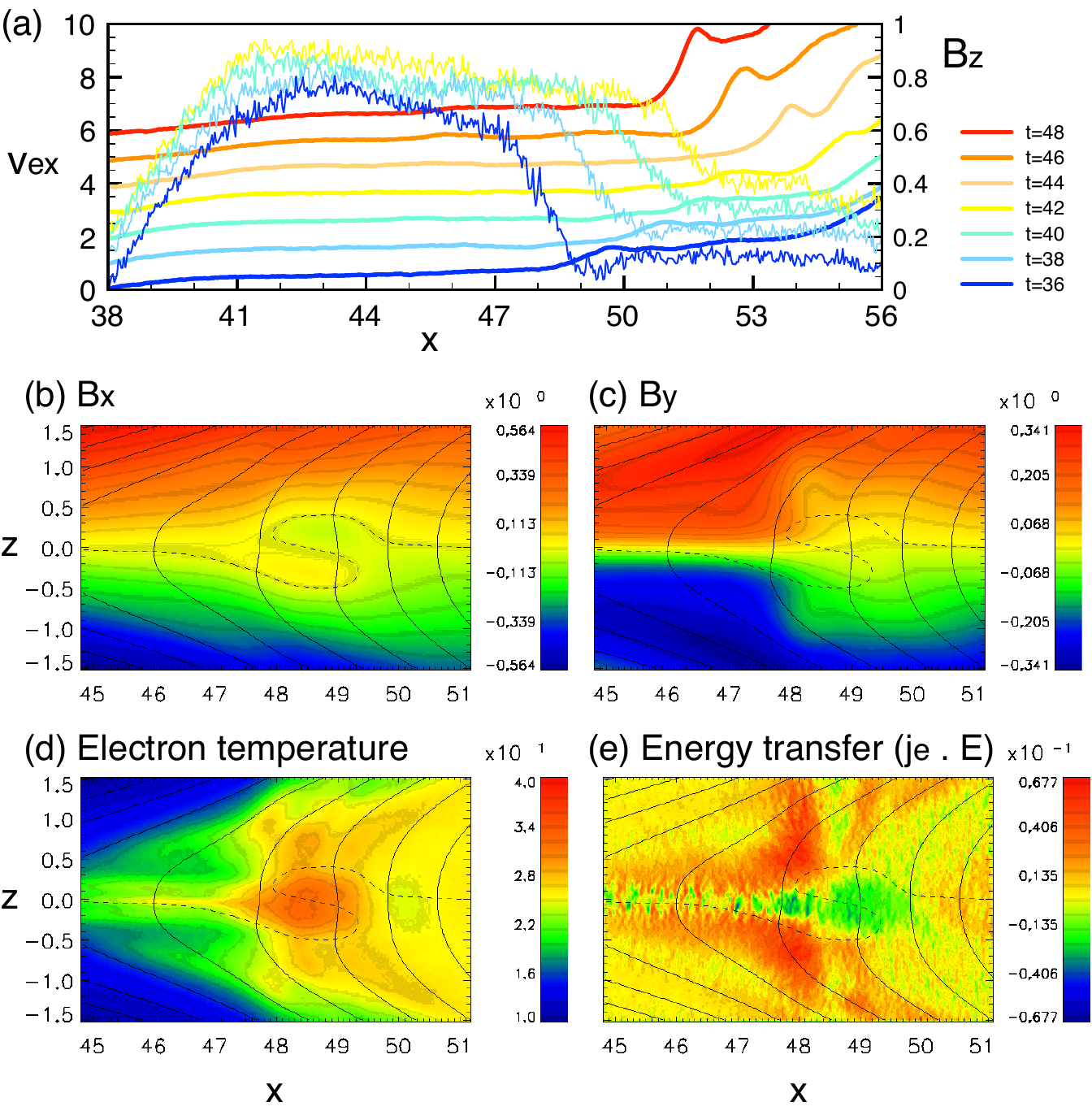}
\fi
\caption{(Color online)
\label{fig:eshock}
(a)
The stack plot of the electron outflow speed $v_{ex}$ (thin lines) and
the reconnected magnetic field $B_z$ (thick lines).
(b-e)
Averaged properties near the electron jet front at $t=35$-$36$:
(b) $B_x$,
(c) the out-of-plane magnetic field $B_y$,
(d) the electron temperature $\frac{1}{3n_e}(p_{xxe}+p_{yye}+p_{zze})$
in units of $m_ec_{Ai}^2$, and
(e)
the electron energy transfer $\vec{j}_e\cdot\vec{E}$ in units of $c_{Ai}B_0 J_0$.
}
\end{center}
\end{figure}

We find a new shock-like structure at the fast electron jet front.
Shown in Figure \ref{fig:eshock}(a) are
the time evolution of the electron outflow velocity ($v_{ex}$) and
the reconnected magnetic field ($B_z$).
At $t=36$, one can see that
$B_z$ becomes twice stronger (from ${\sim}0.08$ to ${\sim}0.16$)
across the transition region, $x\sim 49$.
On the other hand, the fast electron jet suddenly slows down there,
and then the electrons are magnetized further downstream.
We also find that the electron temperature and pressure increase
across the transition region.
These signatures indicate that
it is a shock between the upstream fast electron jet and
the downstream magnetized electron flow.
The discontinuity separates
an unmagnetized upstream flow and
the magnetized downstream flow.
Ions are insensitive to the shock. 
They are not magnetized on both sides, and
the length of the transition region is only one $d_i$,
which is smaller than the ion inertial length based on the local density.
In order to distinguish this from full MHD shocks,
we hereafter refer to it as an ``electron shock.''

As time goes on, the electron shock moves outward.
It reaches at $x\sim 53$ at $t=44$.
After that, the shock propagates backward.
This is the influence of our periodic boundary condition.
After a sufficiently long time,
the back pressure from the downstream region is strong enough to
push the shock front backward,
like a reverse shock reflected by a wall.
The shock becomes stronger than the earlier phase,
as seen in $B_z$ in Figure \ref{fig:eshock}(a).

Shown in Figures \ref{fig:eshock}(b)--\ref{fig:eshock}(e) are
physical quantities near the electron shock region.
Interestingly, $B_x$ changes its sign
in the shock transition region of $48<x<49$:
$B_x<0$ on the upper side and $B_x>0$ on the lower side [Fig. \ref{fig:eshock}(b)],
due to a reverse electron current $j_{ey}<0$.
Unlike the dissipation region, electrons travel in the +$y$-direction there,
because the fast-traveling electrons are just trapped by $B_z$.
In the upstream region of $x<48$,
$B_y$ sharply changes its polarity near the electron jet,
while it is rather uniform outside the jet [Fig. \ref{fig:eshock}(c)].
In the shock-downstream of $x>48$,
$B_y$ suddenly becomes weaker than in the upstream.
The vertical magnetic pressure $\frac{1}{2\mu_0}(B^2_x+B^2_y)$ exhibit
similar profiles.
This is identical to the magnetic cavity (Figure \ref{fig:3D}). 
Note that the vertical component ($B_z$) is compressed across the shock.

We calculate the electron temperature [Fig. \ref{fig:eshock}(d)]
by tracing the diagonal components of the pressure tensor,
\begin{equation}
T_e = \frac{1}{3n_e}(p_{xxe}+p_{yye}+p_{zze}).
\end{equation}
Each component basically shows a similar profile.
One can see that the temperature quickly increases across the shock,
due to the fast jet speed.
The energy transfer from the electromagnetic fields to electrons,
$\vec{j}_e\cdot\vec{E}$, is presented in Figure \ref{fig:eshock}(e).
It is negative $\vec{j}_e\cdot\vec{E} < 0$ around the transition region.
The same signature of $\vec{j}_e\cdot\vec{E} < 0$ was reported
by a previous work
(the ``electron dynamo region'' in Ref.~\onlinecite{sitnov09}). 

These results tell us that
the bulk kinetic energy of the electron jet is transferred to
the electron heat [Fig. \ref{fig:eshock}(d)] and
the magnetic energy [Fig. \ref{fig:eshock}(e)] in the shock downstream. 
Here, the magnetic energy is stored in the compressed reconnected field $B_z$,
because both $B_x$ and $B_y$ are very weak near the neutral line $z=0$.
The strong electron heating also explains the magnetic cavity structure.
Although the electron pressure substantially increases across the shock,
the surrounding tangential fields $(B_x, B_y)$ are not
strong enough to confine electrons in the $z$-direction in the downstream.
The high-pressure electrons expand the structure outwards along $B_z$ in the $\pm z$ directions.
The displacement of the tangential fields lead to the magnetic cavity.

\section{Discussion}
\label{sec:discussion}

Starting with a work by \citet{dau06},
the structures of the electron diffusion region
has been actively discussed
for the past five years.
Earlier investigations focused on
the violation of the ideal frozen-in condition, $\vec{E}' \ne 0$.
This lead to the popular two-scale picture\cite{kari07,shay07},
in which the role of the EDRs was not clearly understood.
In this work, we have explored and reorganized
our understanding of fine reconnection structures
in line of Ref.~\onlinecite{zeni11c}.
We have confirmed that the electron-frame dissipation measure
characterizes the critical region at higher mass ratio of $m_i/m_e=100$.
Since the size of the dissipation region becomes smaller at higher mass ratio,
it seems that the electron physics is responsible for the dissipation region. 
The relevance to the mass ratio further needs to be investigated
in order to extrapolate our results to a realistic ratio of 1836.
In Sec.~\ref{sec:outflow}, we found that
no dissipation takes place in the electron jet region, $D_{\rm mhd}<0$.
Consequently, we don't think it is appropriate to call it
the outer electron ``diffusion'' region or
the outer electron ``dissipation'' region.
As reconnection proceeds,
the nondissipative electron jets are elongated,
while the central dissipation region remains compact.
This is consistent with the recent consensus
that reconnection remains fast \citep{shay07,drake08,wan08d,klimas08}.

From the observational viewpoint,
the contribution from the charge term 
in Eq. \ref{eq:EDR2} is of strong interest,
because the charge density $\rho_c$ will be difficult to probe.
In run 1A, the charge term is an order-of-magnitude smaller than $D_e$, and
it is only localized near the dissipation region [Fig. \ref{fig:35}(f)].
In run 1B with $\omega_{pe}/\Omega_{ce}=2$,
the charge separation appears nonnegligible.
Let us estimate the charge separation around the dissipation region.
We assume that the local plasma density $2n=n_i+n_e$ is roughly uniform.
Electrons are magnetized to the field lines
outside the electron current layer, and
the electron out-of-plane speed is on
an order of the electron Alfv\'{e}n speed $c_{Ae}$.
The Hall electric field $E_z$ is approximated by
\begin{eqnarray}
E_z \approx |v_{ey}| B_0
\sim c_{Ae} B_0 
= c\Big( \frac{\Omega_{ce}}{\omega_{pe}} \Big) B_0
. 
\end{eqnarray}
Gauss's law in the $z$ direction tells us
the charge density in the center,
\begin{eqnarray}
\label{eq:Gauss}
e({n_i-n_e})
\approx
-\frac{\epsilon_0 c}{\delta}
\Big( \frac{\Omega_{ce}}{\omega_{pe}} \Big) B_0
,
\end{eqnarray}
where $\delta$ is the typical thickness of the dissipation region.
Using the result in Sec. \ref{sec:DR},
we assume
\begin{eqnarray}
\label{eq:delta}
\delta \sim \Big(\frac{c}{\omega_{pi}}\Big)
\Big( \frac{m_e}{m_i} \Big)^s
,
\end{eqnarray}
where the index $s$ is between $1/4 \le s \le 1/2$.
Substituting this to Eq. \ref{eq:Gauss},
we obtain
\begin{eqnarray}
\Big| \frac{n_i-n_e}{n_i+n_e} \Big|
\sim
\Big( \frac{m_i}{m_e} \Big)^s
\frac{\omega_{pi}}{c}
\frac{\epsilon_0 c}{2 en}
\Big( \frac{\Omega_{ce}}{\omega_{pe}} \Big) B_0
~~~~~~~~~~~~~~~
\nonumber
\\
\propto
\Big( \frac{m_i}{m_e} \Big)^s
\Big( \frac{\omega_{pi}}{\omega_{pe}} \Big)
\Big( \frac{\Omega_{ce}}{\omega_{pe}} \Big)^2
=
\Big( \frac{m_i}{m_e} \Big)^{s-1/2}
\Big( \frac{\Omega_{ce}}{\omega_{pe}} \Big)^2
.~~
\end{eqnarray}
We see that
the charge separation is controlled
by the $\omega_{pe}/\Omega_{ce}$ parameter
and 
that
it is  much less sensitive to the mass ratio.

In the magnetotail,
the magnetic field is $B_0 \approx 20$ nT in the lobe (upstream region).
The corresponding electron gyro frequency is $f_{ce}=(\Omega_{ce}/2\pi)\approx 560$ Hz.
The plasma sheet density $n_0 \sim 1~{\rm cm}^{-3}$ gives $f_{pe}=(\omega_{pe}/2\pi) \sim 9$ kHz.
Thus we obtain a typical value of $\omega_{pe}/\Omega_{ce} \sim 16$.
In addition, since the upstream plasma occupies the reconnection region,
it is useful to compare the upstream conditions.
While our numerical model assumes that the background density is $0.2n_0$,
the lobe density is substantially lower,
$\mathcal{O}(0.1) - \mathcal{O}(0.01)~{\rm cm}^{-3}$.
In the case of $0.02~{\rm cm}^{-3}$,
since an upstream plasma frequency is modified by a factor of $\sqrt{10}$,
our model with $\omega_{pe}/\Omega_{ce} \sim 5$ will be relevant.
These estimates suggest that
the charge separation effects will be marginally negligible in the magnetotail.
This needs to be verified by further numerical investigations
with realistic lobe densities\citep{bessho10,penny11}.

\begin{figure*}[thp]
\begin{center}
\ifjournal
\includegraphics[width=\textwidth]{f9.eps}
\else
\includegraphics[width=\textwidth]{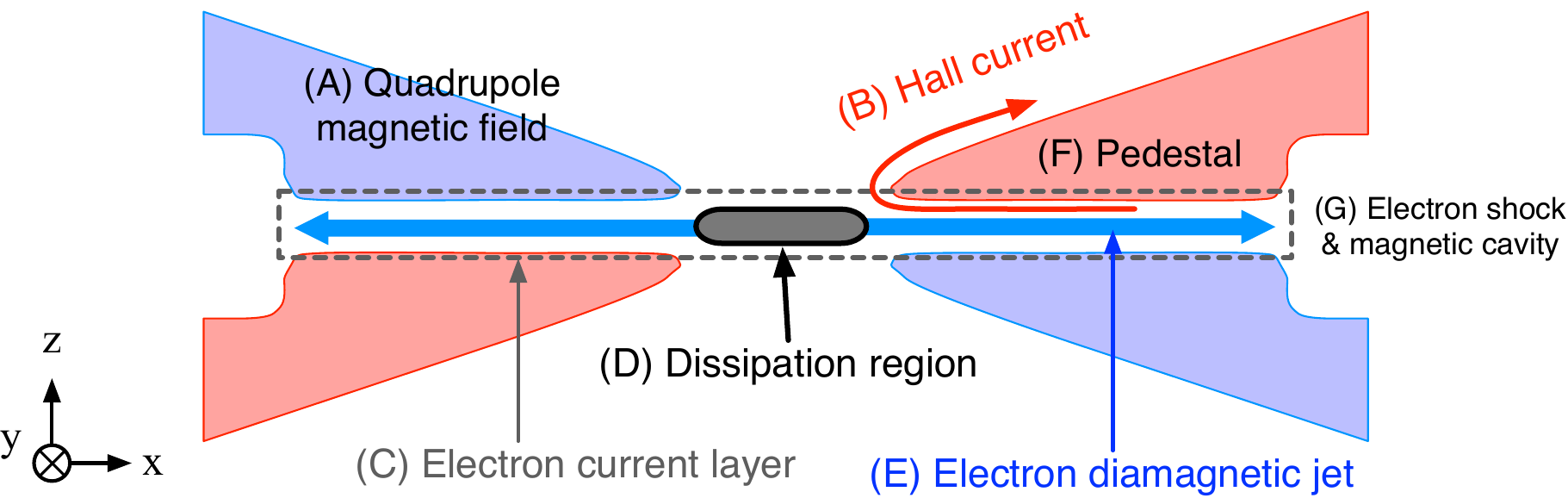}
\fi
\caption{(Color online)
\label{fig:summary}
Our present understanding of Hall reconnection structure:
(A) Quadrupole magnetic field $B_y$ (Refs.~\onlinecite{sonnerup79} and \onlinecite{terasawa83}),
(B) Hall current system (Ref.~\onlinecite{sonnerup79}),
(C) electron current layer (Refs.~\onlinecite{dau06} and \onlinecite{keizo06}),
(D) dissipation region (Ref.~\onlinecite{zeni11c}; Sec. \ref{sec:DR}),
(E) electron diamagnetic jet
(Refs.~\onlinecite{kari07}, \onlinecite{shay07}, \onlinecite{hesse08}; Sec. \ref{sec:outflow}),
(F) pedestal (Ref.~\onlinecite{drake08}), and
(G) electron shock and magnetic cavity (Sec. \ref{sec:eshock}).
}
\end{center}
\end{figure*}

We have found that the fast electron jet 
terminates at the shock, where the electrons become magnetized.
The magnetic cavity structure can be understood as its consequence.
We expect that shock-like signatures are more prominent at higher mass ratio,
because the electron jet speed ${\sim}c_{Ae}$ is much faster than
the typical outflow speed ${\sim}c_{Ai}$.
In a larger system or an open-boundary system\citep{dau06,klimas08},
we expect that the electron shock travels further downstream,
unless
(1) an obstacle in the downstream such as a secondary island,
(2) instabilities of the jet, and
(3) 3D effects
interrupt the shock propagation.
The electron shock may play a role on particle acceleration as well.
For example, \citet{hoshino01} discussed
a two-step scenario of electron acceleration:
pre-acceleration near the reconnection site and
energetization by $\grad${\bf B}/curvature drifts near the flux the flux pile-up region.
Our results suggest that a significant shock-heating
(and possibly shock-drift type acceleration)
takes place between the two acceleration sites. 
The electron shock deserves further investigation
in the context of particle acceleration.

We summarize our understanding in Figure \ref{fig:summary}.
As well known, a Hall reconnection features
the quadrupole out-of-plane magnetic field $B_y$,
which is generated by the Hall current circuit \citep{sonnerup79}.
There is a narrow channel of $E'_y \ne 0$ near the neutral plane \citep{dau06}.
We call this channel the electron current layer,
because the electric current $|j_e|$ is
distinctly stronger than in the other regions.
This current layer is beyond the scope of the fluid theory.
Inside the electron current layer,
there is a single dissipation region
surrounding the reconnection point \citep{zeni11c}.
This region is thinner and longer than the conventional inner region.
The fast, narrow electron jet \citep{kari07,shay07}
travels from the dissipation region.
This is a projection of the diamagnetic current\citep{hesse08} and
a part of the Hall current circuit\citep{sonnerup79}.
As discussed in Section \ref{sec:outflow}, this region is not dissipative.
There is a pedestal region\citep{drake08} outside the electron jet.
The electron jet is terminated by an electron shock and
the magnetic cavity develops there.

The upcoming MMS mission will measure
the field and plasma properties at high resolutions
in near-Earth reconnection sites.
Due to the limited bandwidth between the satellite and the Earth,
it is very important to select the data in high-priority regions.
We have demonstrated that the electron-frame dissipation measure
is a good marker of the most important dissipation region. 
We have further updated our understanding on fine structures
surrounding the dissipation region.
We hope that the measure is helpful to maximize
the scientific profit of the mission.

\begin{acknowledgments}
The authors acknowledge Keizo Fujimoto and Nicolas Aunai for useful comments. 
One of the authors (S.Z.) acknowledges support from
JSPS Postdoctoral Fellowships for Research Abroad.
This work was supported by NASA's MMS mission.
\end{acknowledgments}

\section{Postscript}

This preprint (arXiv:1110.3103v4) contains
all the corrections in our erratum paper\citep{zeni14}.

\end{document}